\documentclass{IEEEtran}
\usepackage{url}
\usepackage{amsmath}
\usepackage{amssymb}
\usepackage{amsthm}
\usepackage{authblk}
\usepackage{upgreek}
\usepackage[ruled,algonl,lined]{algorithm2e}
\usepackage{cite}
\usepackage[dvipsnames]{xcolor}
\usepackage{comment}

\newtheorem*{corollary*}{Corollary}

\usepackage[normalem]{ulem}

\DeclareMathAlphabet{\mathbit}{OML}{cmr}{bx}{it}
\DeclareMathAlphabet{\mathsf}{OT1}{cmss}{m}{n}
\DeclareMathAlphabet{\mathbsf}{OT1}{cmss}{bx}{it}

\newcommand{\diff}{\ensuremath{\:\mathrm{d}}}

\newcommand{\pderiv}[2]{\ensuremath{\frac{\partial#1}{\partial#2}}}
\newcommand{\pderivk}[3]{\ensuremath{\frac{\partial^{#3}#1}{\partial{#2}^{#3}}}}
\newcommand{\Deltarm}{\ensuremath{\mathrm{\Delta}}}

\newlength{\figurewidth}
\newlength{\figureheight}

\usepackage{tikz}
\usetikzlibrary{arrows,decorations,backgrounds,shadows,plotmarks,calc, positioning}
\usetikzlibrary{arrows.meta}

\usepackage{pgfplots}
\pgfplotsset{compat=newest}
\pgfplotsset{plot coordinates/math parser=false}
\pgfplotsset{every axis/.append style={font=\footnotesize}}
\pgfplotsset{
	ylabel right/.style={
		after end axis/.append code={
			\node [rotate=90, anchor=north] at (rel axis cs:1,0.5) {#1};
		}   
	}
}

\usepackage{fancyhdr}
\fancypagestyle{others}{%
	\chead{\scriptsize This paper has been accepted for presentation at the IEEE 17th International Symposium on Wireless Communication Systems (ISWCS) 2021.}
	
	\rhead{\thepage}
	\cfoot{\scriptsize Copyright (c) 2021 IEEE. Personal use is permitted. For any other purposes, permission must be obtained from the IEEE by emailing pubs-permissions@ieee.org.
	}
}
\title{Regular Perturbation and Achievable Rates of Space-Division Multiplexed Optical Channels}
\author{Francisco Javier Garc\'ia-G\'omez and Gerhard Kramer
\IEEEcompsocitemizethanks{
	\IEEEcompsocthanksitem
	Date of current version \today. 
	This work was supported by the German Research Foundation (DFG) under Grants KR 3517/8-1 and 3517/8-2.
	(\emph{Corresponding author: Francisco Javier Garc\'ia-G\'omez.})
	\\ \indent
	The authors are with the Institute for Communications Engineering (LNT), Technical University of Munich, 80333 Munich, Germany (e-mail: javier.garcia@tum.de; gerhard.kramer@tum.de).
	}
	}  

\begin{document}
\maketitle

\begin{abstract}
Regular perturbation is applied to space-division multiplexing (SDM) on optical fibers and motivates a correlated rotation-and-additive noise (CRAN) model. For $S$ spatial modes, or $2S$ complex-alphabet channels, the model has $4S(S+1)$ hidden independent real Gauss-Markov processes, of which $2S$ model phase noise, $2S(2S-1)$ model spatial mode rotation, and $4S$ model additive noise. Achievable information rates of multi-carrier communication are computed by using particle filters. For $S=2$ spatial modes with strong coupling and a 1000 km link, joint processing of the spatial modes gains 0.5 bits/s/Hz/channel in rate and 1.4 dB in power with respect to separate processing of $2S$ complex-alphabet channels without considering CRAN.
\end{abstract}


\thispagestyle{others}
\pagestyle{others}

\section{Introduction}
Space-division multiplexing (SDM) via multi-core and multi-mode transmission considerably increases the capacity of optical fibers~\cite{richardson2013sdm}. Several experiments reaching 10 Pbit/s have been reported, e.g., see~\cite{soma2017petabit, ryf2020transmission, luis2020experimental}. The analysis in~\cite{kramer2015upper,yousefi2015upper} generalizes to SDM and gives a spectral efficiency upper bound of $\log_2(1+\textrm{SNR})$ bits/s/Hz/channel, where SNR is the signal-to-noise ratio without the fiber nonlinearity and ``channel'' refers to a complex-alphabet channel.

Various mismatched models have been developed to compute achievable rates for single and dual polarization (1-pol, 2-pol) transmission~\cite{essiambre_limits,poggiolini2014gn,carena2014egn}. We focus on models based on regular perturbation (RP)~\cite{mecozzi2000analysis, mecozzi2000system,vannucci2002rp,mecozzi2012nonlinear} and logarithmic perturbation (LP)~\cite{Ciaramella-Forestieri-PTL05,Forestieri2005}. A combined RP and LP model is motivated in~\cite{secondini2009crlp}, see also~\cite{secondini2013achievable,secondini2019nonlinearity}.

We follow~\cite{garcia2020mismatched,garcia2021mismatched} and develop an analysis for SDM by using the coupling equations in~\cite{mumtaz2013nonlinear,mecozzi2012coupled}, RP, and an approximation related to LP. The result is a correlated rotation-and-additive noise (CRAN) model with correlated phase noise, spatial mode rotations, and additive noise. We use the model to compute achievable rates for SDM channels.

\emph{Notation:} We use similar notation as in~\cite{garcia2020mismatched} and refer to that paper for details. For instance, we write the Fourier transform of a function $u(t)$ as $\mathcal{F}\left(u(t)\right)$ and the inverse Fourier transform of $U(\Omega)$ as $\mathcal{F}^{-1}\left(U(\Omega)\right)=\mathcal{F}^{-1}\left(U(\Omega)\right)(t)$. For vectors, operators such as $\mathcal{F}$ and $\mathcal{F}^{-1}$ are applied entrywise.

\section{Space-Division Multiplexing}
Each spatial mode has two complex-alphabet channels with the same spatial field distribution but orthogonal polarization. Consider $S$ spatial modes, i.e., $2S$ complex-alphabet channels in total.\footnote{This paper refers to mode-division multiplexing but equation~\eqref{eq:weak_coupling} also applies to multi-core transmission under certain conditions~\cite{mumtaz2013nonlinear}.}
The propagating signal is 
\begin{equation}
\mathbf{u}(z, t)=\left(\begin{matrix}
\mathbf{u}^{[1]}(z, t)^T & \mathbf{u}^{[2]}(z, t)^T & \cdots & \mathbf{u}^{[S]}(z, t)^T
\end{matrix}\right)^T
\end{equation}
where $\mathbf{u}^{[s]}(z,t)=(u^{[s]}(z, t),\,\overline{u}^{[s]}(z, t))^T$ is the vector of signals of the $s$-th spatial mode, $s=1,\dots,S$. The variable $z$ represents distance and $t$ is time. We consider the following two propagation scenarios. The models assume that fiber birefringence changes randomly with $z$~\cite{mumtaz2013nonlinear}.

\emph{Weak Coupling:} the linear coupling among spatial modes is neglected, which is reasonable for multi-core transmission. The propagation equation for spatial mode $s$ is~\cite[Eq. (26)]{mumtaz2013nonlinear}
\begin{align}
& \pderiv{\mathbf{u}^{[s]}}{z} = j\beta_{0}^{[s]}\mathbf{u}^{[s]}-\beta_1^{[s]}\pderiv{\mathbf{u}^{[s]}}{t}-j\frac{\beta_2^{[s]}}{2}\pderivk{\mathbf{u}^{[s]}}{t}{2} + \frac{\mathbf{n}^{[s]}}{\sqrt{g_s(z)}} \nonumber \\
& + j\gamma\left(f_{s,s}g_s(z)\left\|\mathbf{u}^{[s]}\right\|^2+\sum_{r\ne s}f_{s,r}g_r(z)\left\|\mathbf{u}^{[r]}\right\|^2\right)\mathbf{u}^{[s]} 
\label{eq:weak_coupling}
\end{align}
where $\beta_0^{[s]}$, $\beta_1^{[s]}$, and $\beta_2^{[s]}$ are the coefficients of the Taylor expansion of the propagation constant $\beta(\Omega)$ in the angular frequency $\Omega$, averaged over the two polarizations. The noise signals $\mathbf{n}^{[s]}=(n^{[s]}(z, t),\, \overline{n}^{[s]}(z, t))^T$ are independent Wiener processes in $z$ such that, in the absence of nonlinearity ($\gamma=0$),  the accumulated noise in a bandwidth of $\mathcal{B}_{\textrm{ASE}}$ at $z=\mathcal{L}$ has autocorrelation function (ACF) $N_{\textrm{ASE}}\mathcal{B}_{\textrm{ASE}}\textrm{sinc}(\mathcal{B}_{\textrm{ASE}} (t-t'))$. The nonlinear coupling coefficients $\gamma$ and $f_{s,s'}$ are described in~\cite{mumtaz2013nonlinear}. Note that we have absorbed the factors $(8/9)$ and $(4/3)$ from~\cite[Eq. (26)]{mumtaz2013nonlinear} into $f_{s,s'}$. The functions $g_s(z)$ account for attenuation and amplification. Ideal distributed amplification (IDA) has $g_s(z)=1$. %

\emph{Strong Coupling:} the linear coupling between modes is strong, i.e., the vector $\mathbf{u}$ is subject to random unitary transformations that change rapidly with $z$. This has an averaging effect that simplifies~\eqref{eq:weak_coupling} to~\cite[Eq. (38)]{mumtaz2013nonlinear}
\begin{equation}
\pderiv{\mathbf{u}}{z}=-j\frac{\beta_2}{2}\pderivk{\mathbf{u}}{t}{2}+j\gamma\kappa g(z)\left\|\mathbf{u}\right\|^2\mathbf{u}+\frac{\mathbf{n}}{\sqrt{g(z)}}
\label{eq:strong_coupling}
\end{equation}
where $\mathbf{n}(z, t)$ has the same statistics as $(\mathbf{n}^{[1], T}, \ldots, \mathbf{n}^{[S], T})^T$, the average dispersion coefficient is $\beta_2=(\sum_s \beta_{2}^{[s]})/S$, and $\kappa$ is defined in~\cite{mumtaz2013nonlinear}.

We develop a SDM-CRAN model for weak coupling. The model for strong coupling is a special case with the parameters $\beta_0^{[s]}=\beta_1^{[s]}=0$, $\beta_2^{[s]}=\beta_2$ and $f_{s,s'}=\kappa$.

\section{Regular Perturbation}
We expand $\mathbf{u}^{[s]}$ in~\eqref{eq:weak_coupling} in powers of a small perturbation $\gamma$:
\begin{equation}
\mathbf{u}^{[s]}(z, t)=\mathbf{u}_0^{[s]}(z, t)+\gamma\Deltarm\mathbf{u}^{[s]}(z, t)+\mathcal{O}(\gamma^2).
\label{eq:perturbation_expansion}
\end{equation}
The steps are the same as in~\cite{garcia2020mismatched,garcia2021mismatched} and we arrive at

\begin{equation}
\mathbf{u}_0^{[s]}(z, t)=\mathbf{u}_{\textrm{LIN}}^{[s]}(z, t)+\mathbf{u}_{\textrm{ASE}}^{[s]}(z, t)
\label{eq:u_0}
\end{equation}
where $\mathbf{u}_{\textrm{LIN}}^{[s]}(z, t)=\mathcal{D}_z^{[s]} \mathbf{u}^{[s]}(0, t)$
and
\begin{align}
& \Deltarm \mathbf{u}^{[s]}(z,t) \nonumber \\ 
& =j\mathcal{D}_z^{[s]}\left[\int_0^z \mathcal{D}_{-z'}^{[s]}\left(f_{s,s}g_s(z')\left\|\mathbf{u}_0^{[s]}(z', t)\right\|^2 \mathbf{u}_0^{[s]}(z', t) \right. \right. \nonumber \\
& \left. \left. +\sum_{r\ne s} f_{s,r}g_r(z')\left\|\mathbf{u}_0^{[r]}(z', t)\right\|^2 \mathbf{u}_0^{[s]}(z', t) \right)\diff z \right]
\label{eq:u_1}
\end{align}
where
\begin{equation}
\mathcal{D}_z^{[s]}u(t)=\mathcal{F}^{-1}\left(e^{j\left(\beta_0^{[s]}-\beta_1^{[s]}\Omega+\frac{\beta_2^{[s]}}{2}\Omega^2\right)z}\mathcal{F}\left(u(t)\right)\right).
\end{equation}
The effect of $\beta_0^{[s]}$ is a $z$-dependent phase shift, and the effect of $\beta_1^{[s]}$ is a $z$-dependent delay. Similar to~\cite{garcia2021mismatched}, the $2S$ entries of the $S$ noise signals $\mathbf{u}_{\textrm{ASE}}^{[s]}(z, t)$ are independent circularly-symmetric complex Gaussian (CSCG) processes. At the end of the fiber ($z=\mathcal{L}$), their ACF for $\gamma=0$ is $N_{\textrm{ASE}}\,\mathcal{B}_{\textrm{ASE}}\,\textrm{sinc}(\mathcal{B}_{\textrm{ASE}}(t-t'))$.

We use pulse amplitude modulation (PAM) and wavelength division multiplexing (WDM). The indexes $c$ of the WDM channels are in the set
$\left\{c_{\min},\cdots,0,\cdots,c_{\max}\right\}$. The center angular frequency of channel $c$ is $\Omega_c$, and the WDM channel of interest (COI) has $c=0$ and $\Omega_0=0$. The transmitted signal of spatial mode $s$ is
\begin{align}
\mathbf{u}^{[s]}(0, t) & =\sum_{m=-\infty}^{\infty}\left(\begin{matrix}
x_m^{[s]}s(t-mT-\tau_0^{[s]}) \\ \overline{x}_m^{[s]}s(t-mT-\overline{\tau}_0^{[s]})
\end{matrix}\right) \nonumber \\
& +\sum_{c\ne 0} e^{j\Omega_c t}\sum_{k=-\infty}^{\infty} \left(\begin{matrix} b_{c, k}^{[s]}s(t-kT-\tau_c^{[s]}) \\ \overline{b}_{c, k}^{[s]}\overline{s}(t-kT-\overline{\tau}_c^{[s]}) \end{matrix}\right)
\label{eq:wdm}
\end{align}
where $T$ is the symbol period. The base pulse $s(t)$ has most of its energy in the frequency band $|\Omega|\le\pi\mathcal{B}$. The channels may have different delays $\tau_c^{[s]}$ and  $\overline{\tau}_c^{[s]}$. The transmit symbols of the COI ($c=0$) are $x_m^{[s]}$ and $\overline{x}_m^{[s]}$, and the transmit symbols of the interfering frequencies are $b_{c, k}^{[s]}$ and $\overline{b}_{c, k}^{[s]}$, $c\ne0$. We consider symmetric symbol energies and fourth moments:
\begin{align}
\begin{array}{ll}
E=\left\langle |X_m^{[s]}|^2\right\rangle=\left\langle |\overline{X}_m^{[s]}|^2\right\rangle, & \text{all $s$, $m$} \\
E_c=\left\langle |B_{c,k}^{[s]}|^2\right\rangle=\left\langle |\overline{B}_{c,k}^{[s]}|^2\right\rangle, & \text{all $s$, $k$, $c\ne0$} \\
Q_c=\left\langle |B_{c,k}^{[s]}|^4\right\rangle=\left\langle |\overline{B}_{c,k}^{[s]}|^4\right\rangle, & \text{all $s$, $k$, $c\ne0$}.
\end{array}
\end{align}

The receiver applies a band-pass filter $h_{\mathcal{B}}(t)$ followed by linear distortion compensation (LDC) or digital back-propagation (DBP), a matched filter, and a sampler:
\begin{equation}
y_m^{[s]}=\int_{-\infty}^{\infty} s_0^{ *}(t-mT)\left\{\mathcal{D}_{-\mathcal{L}}^{[s]}\left[h_{\mathcal{B}}(t)*u^{[s]}(\mathcal{L}, t)\right]\right\}\diff t.
\label{eq:receiver}
\end{equation}
We focus on $y_m^{[s]}$ to save space. Due to symmetry, the model for $\overline{y}_m^{[s]}$ is obtained by replacing all variables with a ``bar'' by the corresponding variables without a ``bar'', and vice versa. We assume that $h_{\mathcal{B}}(t)*s(t)=s(t)$.

Substituting~\eqref{eq:perturbation_expansion} into~\eqref{eq:receiver}, and using~\eqref{eq:wdm},
we obtain the RP model of the received symbols:
\begin{equation}
y_m^{[s]}=x_m^{[s]}+w_m^{[s]}+\Delta x_m^{[s]}.
\label{eq:y}
\end{equation}
As in~\cite{garcia2021mismatched}, the noise $w_m^{[s]}$ is independent and identically distributed (i.i.d.) complex Gaussian with variance $N_{\textrm{ASE}}$. Also as in~\cite{garcia2021mismatched}, we neglect signal-noise mixing, and the nonlinear interference (NLI) becomes
\begin{align}
\Delta x_m^{[s]} & =j\sum_{s'} \sum_{\substack{n\\ k, k'}} S_{n,k,k'}^{[s,s']}x_{n+m}^{[s]}x_{k+m}^{[s']}x_{k'+m}^{[s'], *} \nonumber \\
& + j\sum_{s'} \sum_{\substack{n\\ k, k'}} \tilde{S}_{n,k,k'}^{[s,s']}x_{n+m}^{[s]}\overline{x}_{k+m}^{[s']}\overline{x}_{k'+m}^{[s'], *} \nonumber \\
& + j\sum_{s'} \sum_{c\ne 0} \sum_{\substack{n\\ k, k'}} C_{c;n,k,k'}^{[s,s']}x_{n+m}^{[s]}b_{c, k+m}^{[s']}b_{c, k'+m}^{[s'], *} \nonumber \\
& + j\sum_{s'} \sum_{c\ne 0} \sum_{\substack{n\\ k, k'}} \tilde{C}_{c;n,k,k'}^{[s,s']}x_{n+m}^{[s]}\overline{b}_{c, k+m}^{[s']}\overline{b}_{k'+m}^{[s'], *} \nonumber \\
& + j\sum_{r\ne s} \sum_{c\ne 0} \sum_{\substack{n\\ k, k'}} D_{c;n,k,k'}^{[s,r]}x_{n+m}^{[r]}b_{c, k+m}^{[s]}b_{c, k'+m}^{[r], *}. \nonumber \\
& + j\sum_{s'} \sum_{c\ne 0} \sum_{\substack{n\\ k, k'}} \tilde{D}_{c;n,k,k'}^{[s,s']}\overline{x}_{n+m}^{[s']}b_{c, k+m}^{[s]}\overline{b}_{c, k'+m}^{[s'], *}. 
\label{eq:Delta_x}
\end{align}
The coefficients in~\eqref{eq:Delta_x} can be written in terms of
\begin{align}
& A_{n,k,k'}^{[s,s']}(t_1, t_2, t_3)=\gamma f_{s,s'}\int_{0}^{\mathcal{L}}\mathrm{d}z\:  g_{s'}(z)\int_{-\infty}^{\infty}\mathrm{d}t\: s^*(z, t) \nonumber \\
& s(z, t-nT-t_1)s(z, t-kT-t_2)s^*(t-k'T-t_3).
\end{align}
We choose $\tau_0^{[s]}=0$. For $r\ne s$, we have
\allowdisplaybreaks
\begin{align}
& S_{n,k,k'}^{[s,s']}=A_{n,k,k'}^{[s,s']}(0, \tau_0^{[s']}, \tau_0^{[s']}) \label{eq:S_nkk} \\
& \tilde{S}_{n,k,k'}^{[s,s']}=A_{n,k,k'}^{[s,s']}(0, \overline{\tau}_0^{[s']}, \overline{\tau}_0^{[s']}) \\
& C_{c;n,k,k'}^{[s,s]}=2A_{n,k,k'}^{[s,s]}(0, \tau_c^{[s]}-\beta_2^{[s]}\Omega_c z, \tau_c^{[s]}-\beta_2^{[s]}\Omega_c z) \\
& C_{c;n,k,k'}^{[s,r]}=A_{n,k,k'}^{[s,r]}(0, \tau_c^{[r]}-\beta_2^{[r]}\Omega_c z, \tau_c^{[r]}-\beta_2^{[r]}\Omega_c z) \\
& \tilde{C}_{c;n,k,k'}^{[s,s']}=A_{n,k,k'}^{[s,s']}(0, \overline{\tau}_c^{[s']}-\beta_2^{[s']}\Omega_c z, \overline{\tau}_c^{[s']}-\beta_2^{[s']}\Omega_c z) \\
& D_{c;n,k,k'}^{[s,r]}=A_{n,k,k'}^{[s,r]}(\tau_0^{[r]}, \tau_c^{[s]}-\beta_2^{[s]}\Omega_c z, \tau_c^{[r]}-\beta_2^{[r]}\Omega_c z) \\
& \tilde{D}_{c;n,k,k'}^{[s,s']}=A_{n,k,k'}^{[s,s']}(\overline{\tau}_0^{[s']}, \tau_c^{[s]}-\beta_2^{[s]}\Omega_c z, \overline{\tau}_c^{[s']}-\beta_2^{[s']}\Omega_c z). \label{eq:D_tilde_nkk}
\end{align}
Using single-polarization DBP on the COI removes the terms with $S_{n,k,k'}^{[s,s]}$ from~\eqref{eq:Delta_x}. Using dual-polarization DBP on the COI removes the terms with $S_{n,k,k'}^{[s,s]}$ or $\tilde{S}_{n,k,k'}^{[s,s]}$. Using multi-mode DBP removes all terms with $S_{n,k,k'}^{[s,s']}$ or $\tilde{S}_{n,k,k'}^{[s,s']}$.

\section{SDM-CRAN Model}\label{sec:2pcpan}
As in~\cite{garcia2021mismatched}, we gather all terms from~\eqref{eq:Delta_x} that depend on the current symbols. Assuming multi-mode DBP, we write
\begin{equation}
\left(\begin{matrix}
\Delta x_m^{[1]} \\
\overline{\Delta x}_m^{[1]} \\
\vdots \\
\Delta x_m^{[S]} \\
\overline{\Delta x}_m^{[S]} \\
\end{matrix}\right)=j\mathbf{J}_m\left(\begin{matrix}
x_m^{[1]} \\
\overline{x}_m^{[1]} \\
\vdots \\
x_m^{[S]} \\
\overline{x}_m^{[S]} \\
\end{matrix}\right)+\left(\begin{matrix}
v_m^{[1]} \\
\overline{v}_m^{[1]} \\
\vdots \\
v_m^{[S]} \\
\overline{v}_m^{[S]} \\
\end{matrix}\right)
\label{eq:Delta_x_vector}
\end{equation}
where $\mathbf{J}_m$ is defined block-wise as
\begin{equation}
\mathbf{J}_m=\left(\begin{matrix}
\mathbf{D}_m^{[1,1]} & \mathbf{B}_m^{[1,2]} & \cdots & \mathbf{B}_m^{[1,S]} \\
\mathbf{B}_m^{[2,1]} & \mathbf{D}_m^{[2,2]} & \ddots & \mathbf{B}_m^{[2,S]} \\
\vdots               & \ddots               & \ddots & \vdots               \\
\mathbf{B}_m^{[S,1]} & \mathbf{B}_m^{[S,2]} & \cdots & \mathbf{D}_m^{[S,S]}
\end{matrix}\right).
\end{equation}
The diagonal blocks (of size $2\times 2$) are
\begin{equation}
\mathbf{D}_m^{[s,s]}=\left(\begin{matrix}
\theta_m^{[s]} & \psi_m^{[s,s]} \\
\overline{\psi}_m^{[s,s]} & \overline{\theta}_m^{[s]}
\end{matrix}\right)
\end{equation}
and the off-diagonal blocks are
\begin{equation}
\mathbf{B}_m^{[s,r]}=\left(\begin{matrix}
\xi_m^{[s,r]} & \psi_m^{[s,r]} \\
\overline{\psi}_m^{[s,r]} & \overline{\xi}_m^{[s,r]}
\end{matrix}\right)
\end{equation}
where, for $r\ne s$, we have 
\begin{align}
\theta_m^{[s]} & = \sum_{s'} \sum_{c\ne 0} \sum_{k, k'} C_{c;0,k,k'}^{[s,s']}b_{c, k+m}^{[s']}b_{c, k'+m}^{[s'], *} \nonumber \\
& + \sum_{s'} \sum_{c\ne 0} \sum_{k, k'} \tilde{C}_{c;0,k,k'}^{[s,s']}\overline{b}_{c, k+m}^{[s']}\overline{b}_{k'+m}^{[s'], *}  \label{eq:theta} \\
\psi_m^{[s,s']} & = \sum_{c\ne 0} \sum_{k, k'} \tilde{D}_{c;0,k,k'}^{[s,s']}b_{c, k+m}^{[s]}\overline{b}_{c, k'+m}^{[s'], *}. \label{eq:psi} \\
\xi_m^{[s,r]} & = \sum_{c\ne 0} \sum_{k, k'} D_{c;0,k,k'}^{[s,r]}b_{c, k+m}^{[s]}b_{c, k'+m}^{[r], *} \label{eq:xi}
\end{align}
and $\overline{\theta}_m^{[s]}$, $\overline{\psi}_m^{[s,s']}$, and $\overline{\xi}_m^{[s, r]}$ are obtained by swapping $b_{c,\ell}^{[i]}$ with $\overline{b}_{c,\ell}^{[i]}$ in~\eqref{eq:theta}-\eqref{eq:xi}, swapping $\tau_c^{[i]}$ with $\overline{\tau}_c^{[i]}$ in~\eqref{eq:S_nkk}-\eqref{eq:D_tilde_nkk}, and shifting all delays such that $\overline{\tau}_0^{[s]}=0$. The residual NLI terms $v_m^{[s]}$ and $\overline{v}_m^{[s]}$ are given by~\eqref{eq:Delta_x} without the two first lines (in the DBP case) and without the summands with $n=0$.

We choose all pulses of the same frequency channel to be synchronized, i.e., we set $\tau_c^{[s]}=\overline{\tau}_c^{[s]}:=\tau_c$ for all $s$ and $c$.\footnote{For multi-carrier transmission, the $2S$ pulses of each subcarrier are synchronized but the pulses of different subcarriers are not.}
We thus have $\tilde{D}_{c;0,k,k'}^{[s,s']}=\tilde{D}_{c;0,k',k}^{[s',s], *}$, ${D}_{c;0,k,k'}^{[s,s']}={D}_{c;0,k',k}^{[s',s], *}$, and
\begin{equation}
\psi_m^{[s,s']}=\overline{\psi}_m^{[s',s], *},\quad \xi_m^{[s,r]}=\xi_m^{[r,s], *},\quad
\overline{\xi}_m^{[s,r]}=\overline{\xi}_m^{[r,s],*}. 
\end{equation}
Since the $\theta_m^{[s]}$ and $\overline{\theta}_m^{[s]}$ are real~\cite{garcia2021mismatched}, the matrix $\mathbf{J}_m$ is Hermitian. Using~\eqref{eq:theta}-\eqref{eq:xi}, it follows that all the diagonal terms $\Theta_m^{[s]}$ and $\overline{\Theta}_m^{[s]}$ are correlated and all other crosscorrelations are 0.

\subsection{Mode Rotation Approximation}
As in~\cite{garcia2021mismatched}, we apply an LP-like approximation
\begin{equation}
\mathbf{I}+j\mathbf{J}_m\approx \exp\left(j\mathbf{J}_m\right).
\label{eq:exp_approx}
\end{equation}
Using~\eqref{eq:Delta_x_vector} and~\eqref{eq:y}, we write the SDM-CRAN model as
\begin{equation}
\mathbf{y}_m=\exp\left({j\mathbf{J}_m}\right)\,\mathbf{x}_m+\mathbf{v}_m+\mathbf{w}_m
\label{eq:simplified_model}
\end{equation}
where $\mathbf{y}_m=(y_m^{[1]}, \overline{y}_m^{[1]},\cdots, y_m^{[S]}, \overline{y}_m^{[S]})^T$, and $\mathbf{x}_m$, $\mathbf{v}_m$, and $\mathbf{w}_m$ are defined similarly. The matrix $\exp\left(j\mathbf{J}_m\right)\in\mathbb{C}^{2S\times 2S}$ is unitary. The components of the residual NLI $\mathbf{v}_m$ are uncorrelated and have memory in the time parameter $m$, see~\cite{garcia2021mismatched}. The statistics of the entries of $\mathbf{J}_m$ can be derived as in~\cite{garcia2021mismatched}. We state results for strong coupling next.

\subsection{Statistics of $\mathbf{J}_m$ for Strong Coupling}\label{sec:statistics}
Strong coupling has $f_{s,s'}=\kappa$, $\beta_2^{[s]}=\beta_2$ and $g_s(z)=g(z)$. Furthermore, all the coefficients of the same frequency channel $c$, except for $C_{c;n,k,k'}^{[s,s]}$, are equal and independent of $s$, $s'$, and $r$. We call these coefficients $X_{c;n,k,k'}$:
\begin{align}
X_{c;n,k,k'} & :=\frac{C_{c;n,k,k'}^{[s,s]}}{2}=C_{c;n,k,k'}^{[s,r]}=\tilde{C}_{c;n,k,k'}^{[s,s']} \nonumber \\
& =D_{c;n,k,k'}^{[s,r]}=\tilde{D}_{c;n,k,k'}^{[s,s']}
\end{align}
for all $s, s'$ and all $r\ne s$. The means of the entries $J_{i,k}[m]$ of the  matrix $\mathbf{J}_m$ in~\eqref{eq:simplified_model} are
\begin{align}
& \left\langle J_{i,i}[m]\right\rangle=(1+2S)\sum_{c\ne 0} E_c \sum_{k}X_{c;0,k,k'} \\
& \left\langle J_{i,k}[m]\right\rangle=0, \quad k\ne i.
\end{align}
The auto- and crosscovariance functions of the $J_{i,k}[m]$ can be written in terms of the following two functions:
\begin{align}
r[\ell] & =\sum_{c\ne 0}(Q_c-E_c^2)\sum_k X_{c;0,k,k}X_{c;0,k-\ell,k-\ell} \nonumber \\
& +\sum_{c\ne 0}E_c^2\sum_{k\ne k'} X_{c;0,k,k'}X_{c;0,k-\ell,k'-\ell}^* \label{eq:r} \\
s[\ell] & =\sum_{c\ne 0}E_c^2\sum_{k, k'} X_{c;0,k,k'}X_{c;0,k-\ell,k'-\ell}^*.
\label{eq:s}
\end{align}
If one defines
\begin{equation}
r_{iki'k'}[\ell]=\left\langle J_{i,k}[m]J_{i',k'}^*[m+\ell]\right\rangle-\left\langle J_{i,k}[m]\right\rangle\left\langle J_{i',k'}^*[m+\ell]\right\rangle
\end{equation}
then the autocovariance functions are
\begin{align}
& r_{iiii}[\ell]=(3+2S)r[\ell] \label{eq:r_iiii} \\
& r_{ikik}[\ell]=s[\ell], \quad k\ne i;
\end{align}
the crosscovariance function of two diagonal elements is
\begin{align}
r_{iikk}[\ell]=(2+2S)r[\ell], \quad k\ne i;
\end{align}
and the other crosscorrelation functions are $0$:
\begin{align}
& r_{iii'k'}[\ell]=0,\quad \ i'\ne k' \\
& r_{iki'k'}=0,\quad (i,k)\ne(i',k')\textrm{ and }(i,k)\ne(k',i')
\end{align}
where we recall that $\mathbf{J}_m$ is Hermitian. If the inputs $B_{c,k}^{[s]}$ and $\overline{B}_{c,k}^{[s]}$ are CSCG, then we have $r[\ell]=s[\ell]$. In the limit of large accumulated dispersion, the approximations introduced in~\cite{dar2013properties} apply (see~\cite{garcia2020mismatched,garcia2021mismatched}):
\begin{align}
& \left\langle J_{i,i}[m] \right\rangle\approx (1+2S)\gamma\kappa\frac{\mathcal{L}}{T}\sum_{c\ne 0}E_c \label{eq:E_theta_approx} \\
& r[\ell]\approx \frac{\gamma^2\kappa^2\mathcal{L}}{T}\sum_{c\ne 0}\frac{Q_c-E_c^2}{\left|\beta_2\Omega_c\right|}\left[1-\frac{|\ell|T}{\left|\beta_2\Omega_c\right|\mathcal{L}}\right]^+  \label{eq:r_approx} \\
& s[\ell]\approx \frac{\gamma^2\kappa^2\mathcal{L}}{T}\sum_{c\ne 0}\frac{E_c^2}{\left|\beta_2\Omega_c\right|}\left[1-\frac{|\ell|T}{\left|\beta_2\Omega_c\right|\mathcal{L}}\right]^+
\label{eq:s_approx}
\end{align}
where one assumes that the contributions of the summands with $X_{c;0,k,k}$ in~\eqref{eq:r} and~\eqref{eq:s} dominate, see~\cite[Fig. 2]{garcia2021mismatched}. For example, in a 2-pol system ($S=1$), the ACF $r_\Theta[\ell]$ of the diagonal elements of $\mathbf{J}_m$ is $5r[\ell]$ (see~\eqref{eq:r_iiii}), with $r[\ell]$ approximately given by~\eqref{eq:r_approx}. This matches~\cite[Eq. (75)]{garcia2021mismatched}.

\section{Simplified SDM-CRAN Model}\label{sec:simplified}
In the following, the receiver removes the means $\left\langle\Theta_{m}^{[s]}\right\rangle\equiv\left\langle J_{2s-1,2s-1}[m]\right\rangle$ and $\left\langle\overline{\Theta}_{m}^{[s]}\right\rangle\equiv\left\langle J_{2s,2s}[m]\right\rangle$ of the phase noise, and we abuse notation and write $J_{i,i}[m]$ for the resulting zero-mean variables. Based on the statistical analysis of Section~\ref{sec:statistics}, we model the entries of the matrix $\mathbf{J}_m$ in~\eqref{eq:simplified_model} as
\begin{align}
& j_{i,i}[m]=2\phi_i[m]+\sum_{i'\ne i}\phi_{i'}[m] \label{eq:J_diagonal} \\
& j_{k,i}[m]=j_{i,k}^*[m] \label{eq:J_offdiag}
\end{align}
where the $\Phi_i[m]$, $i=1,\ldots,2S$, are independent, real, Gaussian, memory-$\mu$ Markov processes generated using the procedure in~\cite[Section IV.B]{garcia2021mismatched} with autocovariance function $r[\ell]$ in~\eqref{eq:r} for $\ell\in\{-\mu, \ldots, \mu\}$. The $J_{i,k}[m]$ for $i<k$ are independent CSCG memory-$\mu$ Markov processes generated using the same procedure with autocovariance function $s[\ell]$ in~\eqref{eq:s} for $\ell\in\{-\mu, \ldots, \mu\}$. The number of hidden independent \textit{real} Markov processes in $\mathbf{J}_m$ is thus $4S^2$, of which $2S$ are the $\Phi_i[m]$.

As in~\cite{garcia2020mismatched,garcia2021mismatched}, we combine the ASE noise and the residual NLI noise into one additive noise term $\mathbf{z}_{m}=\mathbf{v}_{m}+\mathbf{w}_{m}$. The simplified (mismatched) SDM-CRAN model is
\begin{equation}
\mathbf{y}_{m}=\exp(j\mathbf{J}_m)\,\mathbf{x}_{m}+\mathbf{z}_{m}.
\label{eq:sdm_cpan}
\end{equation} 
We model the entries $Z_i[m]$ of $\mathbf{z}_m$ as independent CSCG processes with real ACFs 
\begin{equation}
r_{Z_i}[\ell]:=\left\langle Z_{i}[m]Z_i^*[m+\ell]\right\rangle=N_{\textrm{ASE}}\delta[\ell]+\left\langle V_{i}[m]V_i^*[m+\ell]\right\rangle
\label{eq:Z_ACF}
\end{equation}
where $v_i[m]$ is the $i$-th entry of $\mathbf{v}_m$. The ACF $r_{Z_i}[\ell]$ has short memory and a similar shape as~\cite[Fig. 3]{garcia2020mismatched}.

\section{Numerical Results}\label{sec:results}

\subsection{Estimating Model Parameters}\label{sec:estimation}
We use a training set to estimate the parameters of the model~\eqref{eq:sdm_cpan}. 
Since the matrix $\exp\left(j\mathbf{J}_m\right)$ in~\eqref{eq:sdm_cpan} is unitary,
if we use i.i.d. Gaussian inputs $\mathbf{x}_m$ and neglect the small correlations in $\mathbf{Z}_m$ then the distribution of $\|\mathbf{Y}_m\|^2$ given $\|\mathbf{X}_m\|$ is noncentral chi-squared with $4S$ degrees of freedom, and independent for each $m$. We thus estimate the noise variance $\sigma_Z^2=r_{Z_i}[0]$ as
\begin{align}
& \hat{\sigma}_Z^2=\arg\max_{\sigma^2} \sum_m \nonumber \\ &   \log\left[\frac{e^{-\frac{\|\mathbf{y}_m\|^2+\|\mathbf{x}_m\|^2}{\sigma^2}}}{\sigma^2} \frac{\|\mathbf{y}_m\|^{2S-1}}{\|\mathbf{x}_m\|^{2S-1}} I_{2S-1}\left(\frac{2\|\mathbf{y}_m\|\|\mathbf{x}_m\|}{\sigma^2}\right)\right].
\end{align}
We estimate the mean phase noise $\langle J_{i,i}[m]\rangle$ as in~\cite{garcia2021mismatched}. For the matrix $\mathbf{J}_m$, we assume that all $\Phi_i[m]$ (see~\eqref{eq:J_diagonal}) have the same ACF $\sigma_\Phi^2r[\ell]$, and that all $J_{k,i}[m]$ for $k<i$ have the same ACF $\sigma_J^2 s[\ell]$, with $r[\ell]$ and $s[\ell]$ given by~\eqref{eq:r_approx} and~\eqref{eq:s_approx}. We minimize the mismatched conditional entropy $h_q(\mathbf{Y}|\mathbf{X})$ (obtained using particle filtering~\cite{garcia2021mismatched}) over $\sigma_\Phi^2$ and $\sigma_J^2$. We use the same symmetric, real, three-tap, unit-energy whitening filter on all modes, and minimize $h_q(\mathbf{Y}|\mathbf{X})$ over its free parameter.

\subsection{Achievable Rates}
We simulated the strong coupling system~\eqref{eq:strong_coupling} with $S=2$ spatial modes and 5 WDM channels by using the split-step Fourier method. The system has IDA and the parameters in Table~\ref{tab:parameters}. As in~\cite{garcia2021mismatched}, we compare single- and multi-carrier systems, where the latter system has four subcarriers (4SC) of bandwidth $12.5$ GHz each. All subcarriers have the same power, i.e., there is no frequency-dependent power allocation as in~\cite{garcia2020mismatched, garcia2021mismatched}. The input symbols have a Gaussian density. The channel delays $\tau_c$ are chosen randomly between $-T/2$ and $T/2$. The receiver applies a 50-GHz band-pass filter to isolate the COI, followed by joint DBP on all COI spatial modes and subcarriers, and then matched filtering and sampling. A training set of $24$ sequences of $4092$ symbols ($20$ sequences of $4\times 1023$ symbols in the 4SC system) is used to estimate the SDM-CRAN model parameters.

\begin{table}[tbp]\centering
	\caption{System parameters}
	\label{tab:parameters}
	\begin{tabular}{ccc}
		\hline
		\textbf{Parameter} & \textbf{Symbol} & \textbf{Value} \\
		\hline
		Dispersion coefficient & $\beta_2$ & $-21.7\;\mathrm{ps}^2/\mathrm{km}$ \\
		Nonlinear coefficient & $\gamma$ & $1.27\;\mathrm{W}^{-1}\mathrm{km}^{-1}$  \\
		RX noise spectral density & $N_{\textrm{ASE}}$ & $5.902\cdot {10}^{-18}\;\mathrm{W}/\mathrm{Hz}$  \\ \hline
		Number of spatial modes & $S$ & $2$  \\
		WDM channel indexes & $c_{\min},c_{\max}$ & $-2,2$  \\
		Transmitted pulse shape & $s(t)$ & sinc \\
		Channel bandwidth & $\mathcal{B}$ & $50\;\textrm{GHz}$ \\
		Channel spacing & $\Omega^{(1)}/(2\pi)$ & $50\;\textrm{GHz}$ \\
		\hline
	\end{tabular}
\end{table}

\begin{figure}[tbp]\centering
	\setlength{\figurewidth}{0.85\linewidth}
	\setlength{\figureheight}{0.55\figurewidth}
	\definecolor{mycolor1}{rgb}{0.50000,1.00000,0.00000}%
\definecolor{mycolor2}{rgb}{0.00000,1.00000,0.25000}%
\definecolor{mycolor3}{rgb}{0.50000,0.00000,1.00000}%
\definecolor{mycolor4}{rgb}{0,0.5,0}%
\definecolor{mycolor5}{rgb}{1,0.5,0}
\definecolor{mycolor6}{rgb}{0,0.5,1}
\definecolor{mycolor7}{rgb}{0,0,0.5}
\begin{tikzpicture}

\begin{axis}[%
width=0.976\figurewidth,
height=1.1\figureheight,
at={(0\figurewidth,0\figureheight)},
scale only axis,
xmin=-12,
xmax=-5,
xlabel style={font=\color{white!15!black}},
xlabel={Power per WDM channel, $\mathcal{P}=E/T$ (dBm)},
ymin=7.4,
ymax=8.4,
ytick distance=0.2,
grid,
ylabel style={font=\color{white!15!black}},
ylabel={Rate (bits/s/Hz/channel)},
axis background/.style={fill=white},
legend columns=5,
transpose legend,
legend style={at={(0.5, 1.01)}, anchor=south, legend cell align=left, align=left, draw=white!15!black}
]

\addplot [color=mycolor5, semithick, mark=diamond, mark options={solid, mycolor5, scale=1.5}]
table[row sep=crcr]{%
-12	7.59609696339315\\
-11	7.84859156472405\\
-10	8.05050012689746\\
-9	8.20157912790291\\
-8	8.24000842940592\\
-7.5	8.20369536647538\\
-7	8.13280470388693\\
-6	7.87952193683034\\
-5	7.45850790433265\\
-4	6.98537328261585\\
};
\addlegendentry{SDM-CRAN}

\addplot [color=red, mark=triangle, mark options={solid, red}]
  table[row sep=crcr]{%
-12	7.58389740300176\\
-11	7.81957766813493\\
-10	7.98993960199017\\
-9	8.07271016773356\\
-8	8.02447355182586\\
-7.5	7.93609137342456\\
-7	7.79759106886708\\
-6	7.43206910938989\\
-5	6.94322216775471\\
-4	6.38783974394068\\
};
\addlegendentry{2pCPAN}

\addplot [color=mycolor3, mark=square, mark options={solid, mycolor3}]
  table[row sep=crcr]{%
-12	7.57689954762313\\
-11	7.80430319400606\\
-10	7.95852218937594\\
-9	8.00165705638529\\
-8	7.90880386095974\\
-7.5	7.79770657112702\\
-7	7.63273501014005\\
-6	7.23377996525327\\
-5	6.72017232317908\\
-4	6.15736774441296\\
};
\addlegendentry{1pCPAN}

\addplot [semithick, dashed, color=black, mark=o, mark options={solid, black}]
  table[row sep=crcr]{%
-12	7.49534606680952\\
-11	7.66591615961296\\
-10	7.7111480140219\\
-9	7.64075120357957\\
-8	7.4235606738431\\
-7.5	7.28601769936499\\
-7	7.06629898985108\\
-6	6.59711686645413\\
-5	6.03474523085052\\
-4	5.41675095811901\\
};
\addlegendentry{1p memoryless}

\addplot [semithick, color=black]
table[row sep=crcr]{%
-12	7.74695458877323\\
-11	8.07776545091807\\
-10	8.40885959646594\\
-9	8.74017910952567\\
-8	9.0716778580849\\
-7.5	9.23748265659658\\
-7	9.40331911732814\\
-6	9.73507366456607\\
-5	10.0669182550719\\
-4	10.3988344044816\\
};
\addlegendentry{$\log_2 (1+\textrm{SNR})$}

\addplot [color=mycolor6, semithick, mark=diamond*, mark options={solid, mycolor6, fill=mycolor6, scale=1.5}]
table[row sep=crcr]{%
-12	7.63944365657811\\
-11	7.89489619268218\\
-10	8.11038435624478\\
-9	8.26147829161037\\
-8	8.31309666283293\\
-7.5	8.29547263543749\\
-7	8.23463962447343\\
-6	8.00556367019223\\
-5	7.6319809144704\\
-4	7.11073079321986\\
};
\addlegendentry{SDM-CRAN, 4SC}

\addplot [color=mycolor7, mark=triangle*, mark options={solid, fill=mycolor7, mycolor7}]
  table[row sep=crcr]{%
-12	7.62745186732697\\
-11	7.87011918168512\\
-10	8.06094634482836\\
-9	8.16543367129298\\
-8	8.13310674525856\\
-7.5	8.07333867590699\\
-7	7.96310799959099\\
-6	7.61704569799586\\
-5	7.16019563612216\\
-4	6.59986437116572\\
};
\addlegendentry{2pCPAN, 4SC}

\addplot [color=mycolor4, mark=square*, mark options={solid,fill=mycolor4, mycolor4}]
  table[row sep=crcr]{%
-12	7.6156419301691\\
-11	7.84847103526063\\
-10	8.01900463571647\\
-9	8.0861179496164\\
-8	8.01162703600624\\
-7.5	7.92013195852086\\
-7	7.77322921655232\\
-6	7.38336229117873\\
-5	6.88765034835311\\
-4	6.30683076964075\\
};
\addlegendentry{1pCPAN, 4SC}

\addplot [semithick, color=black, dashed, mark=*, mark options={solid, fill=black, black}]
  table[row sep=crcr]{%
-12	7.54712218902077\\
-11	7.71803060483585\\
-10	7.78611430970355\\
-9	7.75248366199729\\
-8	7.53600511358328\\
-7.5	7.42812811444723\\
-7	7.21375961114647\\
-6	6.7575790517093\\
-5	6.22345033047522\\
-4	5.61775086797744\\
};
\addlegendentry{1p memoryless, 4SC}

\end{axis}

\end{tikzpicture}%
	\caption{Achievable rates for a $1000$-km SDM link with $S=2$ spatial modes, $5$ WDM channels, strong coupling, and the parameters in Table~\ref{tab:parameters}. $\textrm{SNR}=P/(N_{\textrm{ASE}}\mathcal{B})$}
	\label{fig:sdm_rates}
\end{figure}
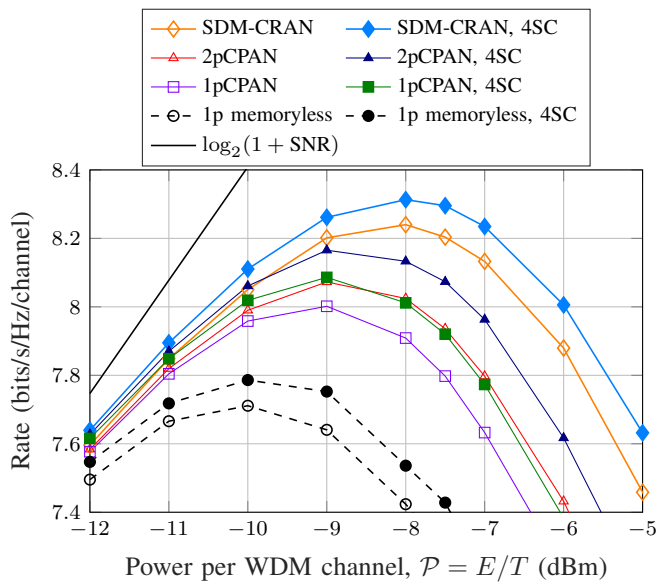

We apply particle filtering on a test set of $120$ sequences ($100$ sequences for 4SC) to compute achievable rates as described in~\cite[Sec. VI]{garcia2021mismatched}. The mismatched output distribution is Gaussian~\cite[Eq. (91)]{garcia2021mismatched}. The results are plotted in Fig.~\ref{fig:sdm_rates} that compares four receiver algorithms with successively more complex processing:
\begin{itemize}
\item separate processing of each of the $2S$ (complex-alphabet) channels using a memoryless mismatched model with i.i.d. phase-and-additive noise, see~\cite[Sec. VIII.A]{garcia2020mismatched};
\item separate 1-pol CPAN (1pCPAN) processing of each of the $2S$ channels with $2S$ particle filters, see~\cite{garcia2020mismatched};
\item separate 2-pol CPAN (2pCPAN) processing of the $S$ spatial modes with $S$ particle filters, see~\cite{garcia2021mismatched}; 
\item joint CRAN processing of all $2S$ channels with one particle filter that uses the mismatched model~\eqref{eq:J_diagonal}-\eqref{eq:Z_ACF}.
\end{itemize}
With respect to memoryless processing, the 1pCPAN model gains 0.29 bits/s/Hz/channel, the 2pCPAN model gains a further 0.07 bits/s/Hz/channel, and the SDM-CRAN model gains another 0.17 bits/s/Hz/channel. 4SC gains between 0.07 and 0.1 bits/s/Hz/channel with respect to single-carrier transmission. The rate gain from the peak of the lowermost curve to the peak of the uppermost curve is approximately 0.6 bits/s/Hz/channel. The rate gain from ``Memoryless'' to ``SDM-CRAN'' for either single- or multi-carrier transmission is 0.5 bits/s/Hz/channel. For 4SC, the power gain from the rate peak with memoryless processing to the SDM-CRAN curve at the same rate (7.79 bits/s/Hz/channel) is 1.4 dB.

\section{Conclusions}
We extended the analysis of the CPAN model~\cite{garcia2020mismatched,garcia2021mismatched} to SDM for weak and strong coupling. The SDM-CRAN model and a multi-carrier scheme were applied to strong coupling to obtain achievable rates for a $S=2$ spatial mode system. The rates are 0.6 bits/s/Hz/channel larger than the rates for a single-carrier system with separate and memoryless processing per complex-alphabet channel.

We remark that computational complexity limits the numerical calculations to a small number $S$ of spatial modes. An important direction for future work is speeding up the calculations and designing simplified receivers that exploit the correlations predicted by the SDM-CRAN model in practical multi-mode systems. An interesting theory question is to extend the LP analyses of \cite{Ciaramella-Forestieri-PTL05,Forestieri2005,secondini2009crlp} to SDM.

\bibliographystyle{IEEEtran}
\bibliography{sdm_cpan.bib}

\end{document}